\documentclass[runningheads]{llncs}
\usepackage{amsmath}
\usepackage{graphicx}
\usepackage{subfigure}
\usepackage{microtype}
\usepackage{caption}
\usepackage{graphicx}
\usepackage{amsmath}
\usepackage{color}
\usepackage{amsfonts}
\usepackage{multirow}
\usepackage{tabu}
\usepackage{booktabs}
\usepackage{marvosym}
\usepackage{ifsym}
\begin{document}
\title{Large-scale Multi-granular Concept Extraction Based on Machine Reading Comprehension\thanks{This work is supported by Science and Technology on Information Systems Engineering Laboratory at the 28th Research Institute of China Electronics Technology Group Corporation, Nanjing Jiangsu, China (No. 05202002), National Key Research and Development Project (No. 2020AAA0109302), Shanghai Science and Technology Innovation Action Plan (No. 19511120400) and Shanghai Municipal Science and Technology Major Project (No. 2021SHZDZX0103). 
}}
\titlerunning{Large-scale Multi-granular CE Based on MRC}
\author{Siyu Yuan\inst{1}\orcidID{0000-0001-8161-6429} \and
Deqing	Yang\inst{1}\textsuperscript{\Letter}\orcidID{0000-0002-1390-3861}\and\\ 
Jiaqing	Liang\inst{2}\orcidID{0000-0003-0670-5602} \and
Jilun	Sun\inst{2}\orcidID{0000-0003-4611-5367}\and\\
Jingyue	Huang\inst{1}\orcidID{0000-0002-3880-6785} \and
Kaiyan	Cao\inst{1}\orcidID{0000-0001-5763-3050} \and
Yanghua	Xiao\inst{2,3}\textsuperscript{\Letter}\orcidID{0000-0001-8403-9591} \and
Rui	Xie\inst{4}\orcidID{0000-0002-1116-7418}}
\authorrunning{Siyu Yuan et al.}
\institute{School of Data Science, Fudan University, Shanghai, China \email{\{yuansy17,yangdeqing,jingyuehuang18,kycao20\}@fudan.edu.cn}\and
School of Computer Science, Fudan University, Shanghai, China 
\email{l.j.q.light@gmail.com}, \email{\{jlsun18,shawyh\}@fudan.edu.cn} 
\and
Fudan-Aishu Cognitive Intelligence Joint Research Center, Shanghai, China
\and
Meituan, Beijing, China
\email{rui.xie@meituan.com}
}
\maketitle              % typeset the header of the contribution
\begin{abstract}
The concepts in knowledge graphs (KGs) enable machines to understand natural language, and thus play an indispensable role in many applications. However, existing KGs have the poor coverage of concepts, especially fine-grained concepts. In order to supply existing KGs with more fine-grained and new concepts, we propose a novel concept extraction framework, namely \emph{MRC-CE}, to extract large-scale multi-granular concepts from the descriptive texts of entities. Specifically, MRC-CE is built with a machine reading comprehension model based on BERT, which can extract more fine-grained concepts with a pointer network. Furthermore, a random forest and rule-based pruning are also adopted to enhance MRC-CE's precision and recall simultaneously. Our experiments evaluated upon multilingual KGs, i.e., English \emph{Probase} and Chinese \emph{CN-DBpedia}, justify MRC-CE's superiority over the state-of-the-art extraction models in KG completion. Particularly, after running MRC-CE for each entity in CN-DBpedia, more than 7,053,900 new concepts (instanceOf relations) are supplied into the KG. The code and datasets have been released at \url{https://github.com/fcihraeipnusnacwh/MRC-CE}.

\end{abstract}

\section{Introduction}
The concepts in knowledge graphs (KGs) \cite{YAGO,DBpedia,CN-DBpedia} enable machines to better understand natural languages and thus play an increasingly significant role in many applications, such as question answering \cite{KBQA}, personalized recommendation \cite{ConceptRec}, commonsense reasoning \cite{CommonSense}, etc. Particularly, fine-grained concepts greatly promote the downstream applications. For example, if entity \emph{Google} has `technology company' and `search engine company' as its concepts, a job recommender system would recommend Google rather than Wal-Mart to a graduate from computer science department based on such fine-grained concepts.

Although there have been a great number of efforts on constructing KGs in recent years, the concepts in existing KGs are still far from being complete, especially for fine-grained concepts. In the widely used Chinese KG \emph{CN-DBpedia} \cite{CN-DBpedia}, there are nearly 17 million entities but only 0.27 million concepts in total, and more than 20\% entities even have no concepts. In general, fine-grained concepts contain more than one modifier, and thus have longer characters (words). However, the average character length of CN-DBpedia concepts is only 3.62, implying that most of them are coarse-grained.

\begin{figure}[!htb]
	%\vspace{-0.2cm}
	\centering
	\includegraphics[width=4in]{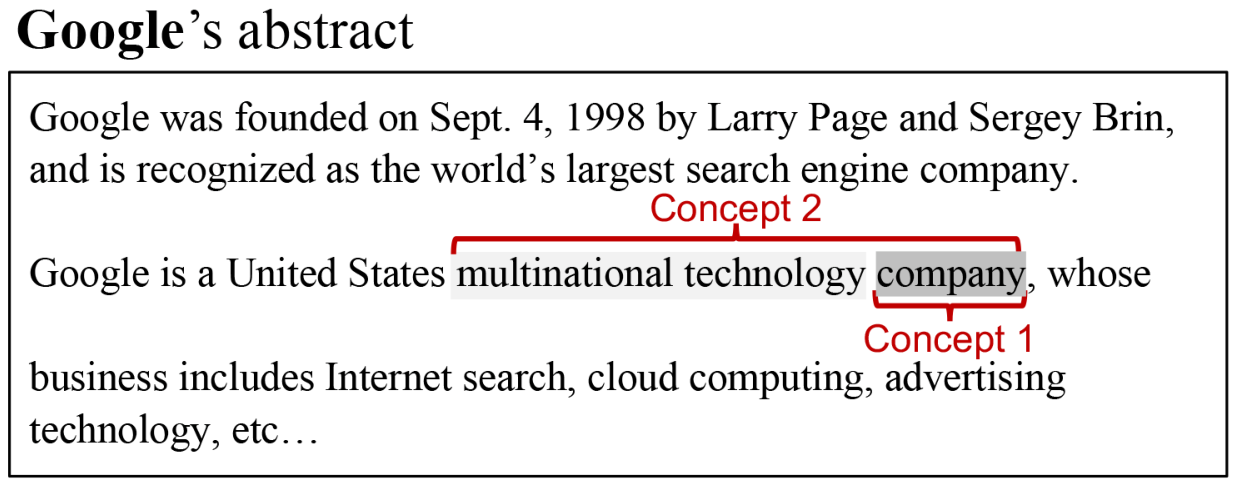} 
	%\vspace{-0.6cm}
	\caption{The abstract text of entity \emph{Google}. The problem of concept overlap in the text is challenging for traditional extraction models to extract multiple concepts from the same span.}
	\label{fig:Google}
	%\vspace{-0.2cm}
\end{figure}

The poor coverage of concepts, especially fine-grained concepts, in the existing KGs is due to their approaches' drawbacks. Most of the existing concept acquisition approaches are based on \emph{generation} or \emph{extraction} from texts. Generation methods often generate coarse-grained concepts from free texts since they are inclined to generate high-frequency words \cite{2019Improving,shvets2020concept}. Extraction methods mainly have three types of models. Pattern-matching models \cite{YAGO,WikiTaxonomy,DBpedia,yao2012evolutionary} focus on extracting concepts from texts based on handcrafted patterns or rules, but the recall of concept extraction (CE) is low due to their limited generalization ability. Classification models \cite{Probase+,xu2018metic,liao2019combining,ji2020fully} identify concepts through classifying a given entity into a predefined concept set based on text information, but can not find new concepts. Sequence labeling models \cite{wei2019novel,nie2020improving} treat the CE problem as a sequence labeling task, but can not handle the problem of \emph{concept overlap}. The concept overlap refers to the phenomenon that a concept term is the subsequence of another concept term. For example, in Fig. \ref{fig:Google}, once a sequence labeling model labels `company' as Concept 1, it would not label `multinational technology company' as a fine-grained concept, since it can not mark a token with different labels. 

Machine reading comprehension (MRC) model can extract the answer from the contextual texts for a proposed question. Inspired by MRC model's excellent extraction capability, we handle CE problem as an MRC task in this paper, and propose a novel CE framework named 
\emph{MRC-CE}, to implement large-scale multi-granular CE from the descriptive texts of given entities. MRC-CE can extract more multi-granular concepts than previous concept extraction models due to the following reasons. Firstly, MRC-CE is built with an MRC model based on BERT \cite{BERT} (BERT-MRC), which can find abundant new concepts and handle the problem of \emph{concept overlap} well with a pointer network \cite{PN}. Secondly, a random forest \cite{ML} and rule-based pruning are adopted to filter the candidate concepts output by BERT-MRC, which enhances MRC-CE's precision and recall simultaneously. Furthermore, MRC-CE has been proven capable of acquiring large-scale multi-granular concepts through our extensive experiments. %since nearly million new concepts have been supplied to CN-DBpedia and the average character length of concepts have been greatly improved by running our MRC-CE models.

The major contributions of this paper can be summarized as follows:

1. To the best of our knowledge, this is the first work to employ MRC model in text-based CE for large-scale KG completion. 

2. Through our experiments, more than 7,053,900 new concepts (instanceOf relations) were extracted by our MRC-CE for completing the large-scale Chinese KG \emph{CN-DBpedia}. Furthermore, the online service of CE based on MRC-CE is released on \url{http://kw.fudan.edu.cn}.

The rest of this paper is organized as follows. Section 2 is the review of related works and Section 3 is the detailed introduction of our framework, respectively. We display our experiment results in Section 4 to justify the effectiveness and rationality of MRC-CE. We also display the CE results of applying our framework on CN-DBpedia and Meituan\footnote{https://www.meituan.com} in Section 5, and at last conclude our work in Section 6.

\section{Related Work}
%\vspace{-0.1cm}
In this section, we review some works related to concept acquisition.
%\vspace{-0.2cm}
\subsection{Ontology Extension}
Ontology \cite{budin2005ontology} extension focuses more on identifying hypernym-hyponym relationships \cite{zhao2018domain,petrucci2018expressive,yilahun2020automatic}. Unlike it, our goal is to acquire more new concepts, especially fine-grained concepts, to complete the existing KGs. Thus, we did not compare MRC-CE with these methods in our experiments.

\subsection{IsA relation and Entity typing}
IsA relation extraction \cite{ruan2016modeling,li2019extracting} is extracting the subsumption (subClassOf) relation between two classes. Entity type aims to classify an entity into a predefined set of types (concepts), such as person, location and organization without new concept recognition \cite{xu2016cross,liang2017transitivity,xu2018metic}. However, in our setting in this paper, the entity is given at first and its candidate concepts are not pre-defined. Thus, our work is different to isA relation extraction and entity typing, and thus the methods of these two tasks were not compared in our experiments.
%\vspace{-0.2cm}

\subsection{Text-based Concept Extraction}
In this paper, we only focus on extracting the concepts already existing in the texts rather than concept generation. Hence we pay attention to extraction models rather than generative models \cite{2019Improving,shvets2020concept}. The existing CE methods can be divided into three categories. 
%\vspace{-0.2cm}
\subsubsection{Pattern-matching Method} 
Pattern-matching methods \cite{YAGO,WikiTaxonomy,DBpedia,Probase,yao2012evolutionary,liang2017probase+,CN-DBpedia,bai2019business} try to extract concepts from free texts based on handcrafted patterns or rules. They can acquire accurate concepts, but only have low recall due to the poor generalization ability. Comparatively, our MRC-CE achieves CE task based on MRC model beyond the limitation of handcrafted patterns, and thus acquires more concepts. 
%\vspace{-0.2cm}

\subsubsection{Learning-based Method}
Classification models transform CE into classification \cite{liang2017transitivity,xu2018metic,liao2019combining,ji2020fully} to determine which concept in a predefined set meets hypernym-hyponym relation with the given entity, but they can not acquire new concepts. Sequence labeling models have been proven effective on extraction tasks \cite{nguyen2019neural,wei2019novel,akbik2019flair,nie2020improving}. Given the extraction feature, sequence labeling models can also extract concepts from the texts describing entities as our MRC-CE. Recently, pre-trained contextual embeddings have been widely used in sequence labeling models \cite{XLM-R,yang2019xlnet,BERT-BiLSTM-CRF,BERT-BiLSTM} to gain good performance, but can not handle the problem of \emph{concept overlap}.
%\vspace{-0.2cm}

\subsubsection{Knowledge-based Method} 
These methods \cite{qiu2019automatic,bai2019business,preum2020emscontext} mainly use the external information from KGs to achieve extraction tasks, resulting in good CE. However, these models have poor generalization ability and only focus on a specific field.
%\vspace{-0.2cm}
\subsection{Machine Reading Comprehension}
MRC \cite{MRC} is a task to seek the answer from contextual texts for a proposed question, which can be categorized into four classes according to answer format, i.e., cloze test, multiple choice, span extraction and free answering. The span extraction \cite{seo2016bidirectional,shen2017reasonet} is most related and similar to our task. The state-of-the-art pre-trained language models \cite{BERT,RoBERTa} are often applied in MRC tasks. More recently, many researchers have employed MRC model in accomplishing other NLP tasks, including nested NER \cite{BERT-MRC} and RE \cite{ERE}. Comparatively, our MRC-CE is the first attempt of employing MRC model in text-based CE.

\section{Methodology}
In this section, we introduce our CE framework in detail.

\subsection{Task Definition}
Our CE task can be formulated as follows. Given an entity $e$ and its relevant descriptive text, denoted as a sequence of words $X=\{x_1, x_2, ..., x_n\}$ where $x_i(1\leq i\leq n)$ is a word token, the CE task aims to find one or multiple spans from $X$ as $e$'s concept(s).

\subsection{Data Construction}
The descriptive text of high quality plays an important role in text-based CE. Since our task scenario is span extraction, the input text should contain the concept(s) of the given entity. We consider the given entity’s abstract in encyclopedia since it is well structured and explicitly mentions the concept(s) of the given entity. In the following introduction, the input text is always referred to the abstract of a given entity. The construction details of our English and Chinese datasets will be introduced in Subsec. \ref{sec:data}.
%To construct the Chinese dataset to evaluate MRC-CE, we randomly selected adequate amount of entities along with their concepts from CN-DBpedia, and took the concepts existing in their abstract texts as the real labels. To construct the English dataset, we also randomly selected adequate amount of entities and their hypernyms (concepts) from Probase \cite{liang2017probase+}, and took the concepts existing in these entities' abstract texts (obtained from Wikipedia) as the real labels.

\subsection{Summary}
\begin{figure*}[!htb]
	%\vspace{-0.2cm}
	\centering
	\includegraphics[width=1.1\columnwidth]{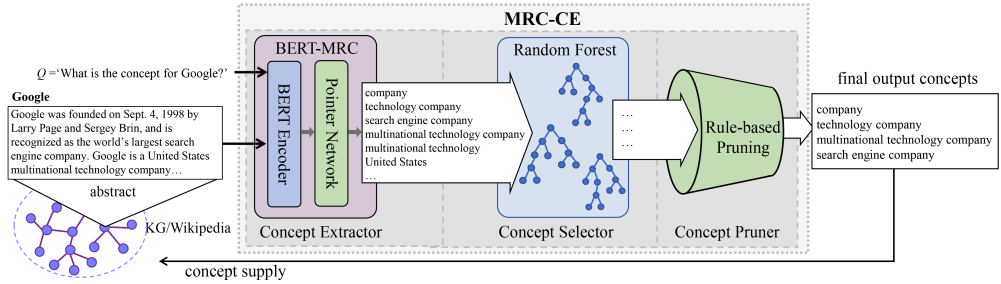} 
	%\vspace{-0.4cm}
	\caption{The pipeline of our MRC-CE framework.}
	\label{fig:pipe}
	%\vspace{-0.1cm}
\end{figure*}

To employ MRC model for CE, we need to construct appropriate questions towards which the spans in the text are extracted. Since our target is to extract concepts from the abstract of a given entity, we set the question $Q$ as `What is the concept for [entity]?'. The pipeline of our MRC-CE is displayed in Fig. \ref{fig:pipe}, which can be divided into three modules, i.e., the concept extractor, concept selector and concept pruner. The first module is BERT-MRC built with a pointer network \cite{PN}, which receives the input text and extracts candidate multi-granular concepts from the text. The second module adopts a random forest \cite{ML} to select the desirable concepts from the candidates output by BERT-MRC. The third module prunes away the unsatisfactory concepts from the second module's outputs based on some pruning rules. %Next, we elaborate these modules.

\subsection{MRC-based Concept Extractor}
The BERT-MRC in our framework is built with a BERT \cite{BERT} encoder and a pointer network to generate candidate spans along with their corresponding confidence scores of being concepts. We took each entity’s concepts from the KG which also exist in the entity’s abstract text as the real labels of one training sample. 

\subsubsection{BERT Encoding}
In this module, BERT is used as an encoder layer to generate an embedding for each input token, based on which the subsequent pointer network predicts the confidence score for each candidate extracted span. At first, the tokens of input $X$ and the tokens of input $X$ and the question $Q$ are concatenated together as follows, to constitute the input of BERT encoder.
$$
\{[CLS]q_1, q_2, ..., q_m[SEP]x_1, x_2, ..., x_n\}
$$
where [CLS] and [SEP] are special tokens, and $q_i(1\leq i\leq m)$ is a token of $Q$. Then, BERT encoder outputs the text token embedding matrix $\mathbf{E}\in\mathbb{R}^{n\times d}$, where $d$ is the embedding dimension.

\subsubsection{Generating Candidate Concepts by Pointer Network}
\begin{figure}[!htb]
	%\vspace{-0.2cm}
	\centering
	\includegraphics[width=0.7\columnwidth]{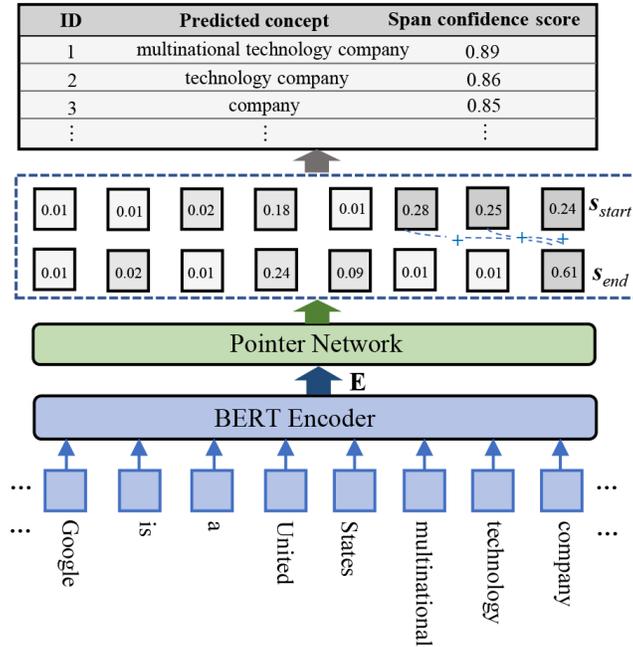} 
	%\vspace{-0.2cm}
	\caption{The structure of BERT-MRC.}
	\label{fig:BERT-MRC}
	%\vspace{-0.2cm}
\end{figure}

With the token embeddings in $\mathbf{E}$, a pointer network is built to predict the probability of a token being the start position and end position of the answer, through a fully connected layer after the BERT encoder. Then, the confidence score of each span can be calculated as the sum of the probability of its start token and end token. One span can be output repeatedly as the same subsequence of multiple extracted concepts through an appropriate selection threshold. This strategy enables extracting multi-granular concepts. For example, in Fig. \ref{fig:BERT-MRC}, `company' is extracted for multiple times corresponding to three concepts of different granularity, if the confidence score threshold is set to 0.85.

Specifically, we use $\mathbf{p}^{start},\mathbf{p}^{end}\in \mathbb{R}^n$ to denote the vectors storing each token's probability of being the start position and end position, respectively. They are calculated based on $\mathbf{E}$ as
\begin{equation}\label{eq:s1}
%\vspace{-0.2cm}
[\mathbf{p}^{start};\mathbf{p}^{end}] = softmax(\mathbf{E}\mathbf{W}+\mathbf{B})
\end{equation}
where $\mathbf{W},\mathbf{B}\in \mathbb{R}^{n\times 2}$ are both trainable parameters. 
Given a span with $x_i$ and $x_j$ as the start token and the end token, respectively, its confidence score ${cs}_{ij}\in \mathbb{R}$ can be calculated as
\begin{equation}\label{eq:s2}
%\vspace{-0.2cm}
cs_{ij} =p^{start}_i+p^{end}_j
\end{equation}
Then, BERT-MRC generates a ranking list of candidate concepts (spans) along with confidence scores, and outputs the concepts with the confidence scores larger than the selection threshold.

\subsubsection{BERT-MRC Loss}
We adopt cross entropy function $CrossEntropy(\cdot)$ as the loss function of BERT-MRC. Specifically, suppose the set $Y_{start}\in \mathbb{R}^n$ (or $Y_{end}\in \mathbb{R}^n$) contains the real label of an input token $x_i(1\leq i\leq n)$ being the start (or end) position of a concept. And $Y_{span}\in \mathbb{R}^{C^2_n}$ contains the real label of a span $x_i x_j$ where $x_i$ and $x_j$ are the start position and end position of a concept, respectively. Then, we have the following three losses for the predictions of the three situations: 
\begin{equation}\label{eq:loss}
\begin{split}
\mathcal{L}_{start}=CrossEntropy(\mathbf{p}^{start},Y_{start})\\
\mathcal{L}_{end}=CrossEntropy(\mathbf{p}^{end},Y_{end})\\
\mathcal{L}_{span}=CrossEntropy(\mathbf{cs},Y_{span})
\end{split}
\end{equation}
Then, the comprehensive loss of training BERT-MRC is 
\begin{equation}\label{eq:loss1}
\mathcal{L}=\alpha \mathcal{L}_{start}+\beta\mathcal{L}_{end}+(1-\alpha-\beta)\mathcal{L}_{span}
\end{equation}
where $\alpha, \beta\in(0,1)$ are control parameters. We use Adam \cite{Adam} to optimize the loss.

\subsection{Selecting Concepts by Random Forest}
The second module (concept selector) is built to select desirable concepts from the candidate spans extracted by BERT-MRC. We argue that it is unsatisfactory to output the concepts by choosing a specific threshold to directly truncate BERT-MRC's output ranking list. If we adopt a relatively big threshold, we can get more accurate concepts but may lose some correct concepts. If recall is preferred, precision might be hurt. To achieve a better trade-off between precision and recall, we adopt a classifer to predict whether a candidate extracted by BERT-MRC deserves being reserved. Such concept selector filters out the desirable concepts in terms of the significant features rather than the span confidence score, and thus improves the performance of concept filtering.

Specifically, we feed the classifer with the following features. At first, we adopt the span confidence score (feature A), start position probability (feature B) and end position probability (feature C) of each candidate span, which are obtained from BERT-MRC. Furthermore, we use another two features. Feature D is whether the current candidate span is an existing concept in the KG. In addition, the most important objective of MRC-CE is to handle the problem of concept overlap for fine-grained CE. Therefore, we consider another feature E whether the current span contains other candidate span(s). 

Since there are only five simple features, we do not need to choose deep models. For the traditional classification model, random forest\cite{ML} has better accuracy than many other models, since it adopts ensemble learning. Therefore, we adopt random forest as the classifier. 

We did not directly use the concepts existing in the KG as the training samples’ real labels of random forest given that many correct concepts not existing in the KG can be extracted by BERT-MRC. Thus, we invited some volunteers to label the training samples manually. Specifically, we randomly sampled 1,000 results from BERT-MRC's outputs and requested the volunteers to label whether the result is a correct concept of the corresponding entity. Then, we saved the features and the labels of the results to constitute the training samples. The significance values of the five features obtained through training the random forest towards the Chinese corpus (CN-DBpedia) are listed in Table \ref{tb:ch_feature}. From the table, we find that except for feature A, feature D is the most significant feature to select desirable concepts. It implies that referring to the existing concepts in KGs is very crucial to extract new concepts.
\begin{table}[!htb]
	%\vspace{-0.2cm}
	\centering
	%\small
	\caption{The significance values of the five features in MRC-CE's random forest for Chinese KG \emph{CN-DBpedia}.}  \label{tb:ch_feature}
	%\vspace{-0.2cm}
	\begin{tabular}{|c| c| c| c| c| c|}
		\hline
		Feature   & A &B & C & D & E \\
		\hline
		Significance  &  0.34 & 0.20 & 0.16 & 0.25 & 0.05  \\  
		\hline
	\end{tabular}
	%\vspace{-0.2cm}
\end{table}

\subsection{Rule-based Concept Pruning}
There are still some wrong concepts being reserved after above concept selection. The errors can be roughly categorized into three classes as below. 

1. The extracted concepts are semantically exclusive with each other. For example, both `president' and `vice president' could be extracted by BERT-MRC, but they are mutually exclusive in terms of conceptualization. Obviously, a president cannot be a vice president simultaneously. 

2. The modifier of a concept is mistakenly extracted as another concept. For example, `railway' and `station' are both extracted as concepts from the span `railway station'. 

3. A correct concept is mistakenly mixed with a functional word. For example, both `is ancient costume drama' and `in high school' are wrong concepts, where `is' and `in' are the redundant tokens.

In this step, some pruning strategies are executed based on explicitly rules, since above errors can be easily recognized according to some pre-defined patterns. For example, for the class that the extracted concepts are mutually exclusive in semantics, we set our pruning rule case by case rather than setting a general rule.

\section{Experiments}
In this section, in order to justify our framework's CE capability for KG completion, we evaluate its performance upon a Chinese KG \emph{CN-DBpedia} and an English KG \emph{Probase}.

\subsection{Datasets}\label{sec:data}
\paragraph{Chinese Dataset}
We obtained the latest data of CN-DBpedia from its open websites\footnote{\url{http://kw.fudan.edu.cn/cndbpedia}} and randomly selected 100,000 entities along with their concepts and abstract texts, as the sample pool. Then, we randomly selected 500 samples from the sample pool as the test set, and the rest samples were divided as the training and validation set according to 9:1, to learn the models. For BERT-MRC, we took each entity’s concepts from the KG which also exist in the entity’s abstract text as the real labels of one training sample. For training the random forest, we invited some volunteers to label whether the candidate concepts from BERT-MRC are the given entity’s concepts since many correct concepts not existing in the KG can be extracted by BERT-MRC. When evaluating the models with the test samples, we invited some volunteers to assess whether the concepts extracted by the models are correct, given that many new concepts not existing in the KG may be extracted.
%Therefore, we do not need inferring their real concepts from KB. 
%Each training sample is constituted by one entity's abstract text and its real concepts existing in the abstract text. 

\paragraph{English Dataset}
Probase\footnote{\url{https://www.microsoft.com/en-us/research/project/probase/}} has the hypernym-hyponym relations between concepts and entities, but no abstract texts of entities. Hence, we first fetched the entities along with their concepts from Probase, and then crawled entities' abstract texts from Wikipedia\footnote{\url{https://www.wikipedia.org/}}. Particularly, we totally sampled 50,000 entities with their concepts and abstract texts. Then, we constructed the training, validation and test set the same as the Chinese dataset.

\subsection{Baselines}\label{sec:base}
We compared our MRC-CE with the following five state-of-the-art models to justify MRC-CE's performance. Please note that, the models we selected are NER and Open IE models since NER and Open IE are extraction tasks which are mostly meet our task scenario. Besides, We also compare the pattern matching approaches.

We did not compare the methods for ontology extension and the generation models, since both of them do not meet our task scenario, i.e., text-based CE. The entity typing models and classification models were also not included because they can not meet the objective to complete existing KGs with new concepts. The knowledge-based methods were also excluded since most of them are only applicable to specific fields. 

\paragraph{Hearst Patterns \cite{2018Hearst}}
Hearst patterns are the basic and popular rules for extracting concepts from free texts. We totally designed 5 Hearst patterns to achieve concept extraction, which are listed in Table \ref{tb:pattern} \footnote{We translate Chinese patterns for CN-DBpedia into English}. We allowed leading and following noun phrases to match limited modifiers to extract fine-grained concepts.
\begin{table}[!htb]
	%\vspace{-0.2cm}
	\centering
	%\small
	\caption{The full set of Hearst patterns for \emph{CN-DBpedia} and \emph{Probase}.}  \label{tb:pattern}
	%\vspace{-0.2cm}
	
	\begin{tabular}{c c}
		\toprule
		\multirow{5}{*}{\textbf{CN-DBpedia}}   & X is Y  \\
		& X is a type/a of Y...  \\
		& X is one of Y   \\
		& X belongs to Y  \\
		& Y is located/founded/ in...  \\
		\midrule
		\multirow{5}{*}{\textbf{Probase}}  &  X is a Y   that/which/who  \\
		& X is one of Y  \\
		& X is a member/part/form... of Y  \\
		& X refers to Y  \\
		& As Y, X is ...  \\
		\bottomrule
	\end{tabular}

\end{table}

\paragraph{XLNet-NER \cite{yang2019xlnet}}
With the capability of modeling bi-directional contexts, XLNet demonstrates an excellent sequence labeling model in many NLP tasks, such as NER and Open IE, which is also applicable to our task. %Due to the extraction feature, XLNet-NER can be employed in the concept extraction as our baseline.

\paragraph{FLAIR \cite{akbik2019flair}}
FLAIR is a novel NLP framework combining different word and document embeddings to achieve excellent results. FLAIR can be employed in our CE task since it can extract spans from the text.

\paragraph{KVMN \cite{nie2020improving}}
As a sequence labeling model, KVMN was proposed to handle NER by leveraging different types of syntactic information through attentive ensemble. %Such model can be transferred to our CE scenario easily since it belongs to sequence labeling model.

\paragraph{XLM-R \cite{XLM-R}}
It is a transformer-based multilingual masked language model incorporating XLM \cite{XLM} and RoBERTa \cite{RoBERTa}. It can achieve CE task since it is also a sequential labeling model.

\paragraph{BERT-BiLSTM-CRF \cite{BERT-BiLSTM-CRF,BERT-BiLSTM}}
It is an advanced version of BERT built with BiLSTM and CRF. This language model is used to obtain optimal token embeddings, based on which the downstream tasks such as NER and Open IE, CE are well achieved.

\subsection{Experiment Settings}
In MRC-CE, we adopted a BERT-base with 12 layers, 12 self-attention heads and H = 768 as BERT-MRC's encoder. The training settings are: batch size = 4, learning rate = 3e-5, dropout rate = 0.1 and training epoch = 2. In addition, we set $\alpha=0.3, \beta=0.25$ in Eq. \ref{eq:loss1} based on parameter tuning. Besides, we chose 50 as the tree number of the random forest.

\subsection{Experiment Results}
%We randomly selected 500 samples from both the Chinese dataset and the English dataset as the test sets, to evaluate each model's CE performance. We further invited some volunteers to assess whether the new concepts extracted by the models are correct. %We display the results of CE comparisons which justify our MRC-CE's superiority over the baselines.

\subsubsection{CE Performance Comparison}
%At first, the absolute number need to be considered since we should ensure whether our model can extract more new and fine-grained concepts than the baselines. Therefore, we compared our MRC-CE with the baselines in terms of absolute numbers. 
We refer to an extracted concept of a given entity that already exists in the KG as an \emph{EC} (existing concept), and refer to an extracted correct concept of a given entity that does not exist in the KG as a \emph{NC} (new concept). NCs are more significant than ECs given that the primary objective of our work is KG completion, i.e., supplying new instanceOf relations. Please note that a NC in fact corresponds to a new instanceOf relation to a given entity rather than being a unique new concept, because the concept linked by this new instanceOf relation may already exist in the KG as another entity's concept.

All models' CE results of the Chinese dataset and the English dataset are listed in Table \ref{tb:result0}, where EC \# and NC \# are the number of existing concepts and new concepts (instanceOf relations) of the entities, respectively. From the table we find that, compared with other models, MRC-CE extracts more NCs. As we claimed before, our framework is capable of extracting fine-grained or long-tail concepts from the texts. To justify it, we counted the average number of characters (for Chinese) or words (for English) constituting ECs (EC length) and NCs (NC length). The larger NC length is, the more possible the NC is to be fine-grained/long-tail. Although Hearst Patterns's NC length is larger than MRC-CE's in the Chinese dataset, it ignores some coarse-grained concepts and thus it can not achieve multi-granular concept extraction. For example, as shown in Table \ref{tb:case}, Hearst Patterns only regards `the railway station of JR East Japan' as a whole concept, whereas MRC-CE can extract `station', `railway station' and `the railway station of JR East Japan' simultaneously from the same span. %Therefore, our MRC-CE is absolutely superior to the baselines in the absolute numbers comparison.

%\paragraph{Comparison In Relative Statistic}
%Furthermore, we also should calculate the relative statistic to realize all-round comparison. We have explained that BERT-MRC can extract overlapped concepts with the help of its in-built pointer network. 

\begin{table*}[!htb]
%\vspace{-0.2cm}
%\small
\begin{center}
\caption{CE performance comparisons of 500 test samples.}\label{tb:result0}
%\vspace{-0.3cm}
\resizebox{\textwidth}{!}
{
    \begin{tabular}
{|c|c|c|c|c|c|c|c|c|c|}
\hline

\hline
\multirow{2}{*}{\textbf{Dataset}} & \multirow{2}{*}{\textbf{Model}} & \multirow{2}{*}{\textbf{EC \#}} &  \multirow{2}{*}{\textbf{NC \#}} &  \textbf{EC} & \textbf{NC} & \textbf{OC} &  \multirow{2}{*}{\textbf{Prec.}} & \multirow{2}{*}{\textbf{R-Recall}} & \multirow{2}{*}{\textbf{R-F1}}\\
& & & &  \textbf{length} & \textbf{Length} & \textbf{ratio} & & &\\
\hline

\hline
& \footnotesize{Hearst} & \multirow{2}{*}{158} &\multirow{2}{*}{222} & \multirow{6}{*}{2.63}& \multirow{2}{*}{\textbf{7.1}} & \multirow{2}{*}{NA} & \multirow{2}{*}{95.24\%} & \multirow{2}{*}{27.24\%} & \multirow{2}{*}{42.36\%} \\
& \footnotesize{Patterns} & & & & &  & & &\\
\cline{2-4} \cline{6-10}
& \footnotesize{XLNet-NER} & 391 & 46 & & 2.61 & NA & 89.92\% & 5.64\% & 10.62\% \\
\cline{2-4} \cline{6-10}
& \footnotesize{FLAIR} & 405 & 63 & & 3.11 & NA & \textbf{95.51\%} & 7.73\% & 14.30\% \\
\cline{2-4} \cline{6-10}
\textbf{CN-}& \footnotesize{KVMN} & 247 & 253 & & 4.03 & NA & 64.27\% & 31.04\%& 41.86\%\\
\cline{2-4} \cline{6-10}
\textbf{DBpedia}& \footnotesize{XLM-R} & 89 & 250 & & 5.35 & NA & 76.35\% & 30.67\% & 43.77\% \\
\cline{2-4} \cline{6-10} 
& \footnotesize{BERT-} & \multirow{2}{*}{411} &\multirow{2}{*}{25} & & \multirow{2}{*}{4.32} & \multirow{2}{*}{NA} & \multirow{2}{*}{88.26\%} & \multirow{2}{*}{3.07\%} & \multirow{2}{*}{5.93\%} \\
& \footnotesize{BiLSTM-CRF} & & & & &  & & &\\
\cline{2-4} \cline{6-10}

& \footnotesize{MRC-CE} & \textbf{519} & \textbf{323} & &4.91 & \textbf{29.35\%} &  {92.22\%} & \textbf{39.63\%} &  \textbf{55.44\%}\\
\hline

\hline
& \footnotesize{Hearst} & \multirow{2}{*}{308} &\multirow{2}{*}{402} & \multirow{6}{*}{1.25}& \multirow{2}{*}{2.04} & \multirow{2}{*}{NA} & \multirow{2}{*}{\textbf{95.56\%}} & \multirow{2}{*}{20.18\%} & \multirow{2}{*}{33.32\%} \\
& \footnotesize{Patterns} & & & & &  & & &\\
\cline{2-4} \cline{6-10}
& \footnotesize{XLNet-NER} & 534 & 398 & & 1.42 & NA & 92.00\% & 19.98\% & 32.83\%\\
\cline{2-4} \cline{6-10} 
& \footnotesize{FLAIR} & 307 & 141 & & 1.67 & NA & 84.69\% & 7.08\% & 13.06\%\\
\cline{2-4} \cline{6-10} 
\textbf{Probase}& \footnotesize{KVMN} & 186 & 404 & & 1.96 & NA & 47.50\%& 20.28\%&28.43\%\\
\cline{2-4} \cline{6-10} 
& \footnotesize{XLM-R} & \textbf{672} & 322 &  & 1.48 & NA & 81.74\% & 16.16\% & 26.99\%\\
\cline{2-4} \cline{6-10} 
& \footnotesize{BERT-} & \multirow{2}{*}{191} &\multirow{2}{*}{154} & & \multirow{2}{*}{1.68} & \multirow{2}{*}{NA} & \multirow{2}{*}{81.18\%}&  \multirow{2}{*}{7.73\%} & \multirow{2}{*}{14.12\%}\\
& \footnotesize{BiLSTM-CRF} & & & & & &  & &\\
\cline{2-4} \cline{6-10} 

& \footnotesize{MRC-CE} & 636 & \textbf{626} & & \textbf{2.31} & \textbf{36.82\%} & {90.08\%} & \textbf{31.38\%} & \textbf{46.54\%}\\
\hline

\hline
\end{tabular}
}
\end{center}
%\vspace{-0.4cm}
\end{table*}

To prove MRC-CE's capability of extracting overlapped concepts, we further recorded the ratio of overlapped concepts (one is another one's subsequence) to all extracted concepts, denoted as OC ratio in Table \ref{tb:result0}. The precision (Prec.) is the ratio of the correct concepts assessed by the volunteers to all concepts extracted by the model. FLAIR obtains the highest Prec. mainly due to its high precision of ECs, but it is howbeit meaningless to KG completion. 
The denominator of recall is the number of all new (correct) concepts in the input text. Since it is difficult to know all new concepts in the input text except for the costly human assessment, we report the relative recall (R-Recall) to measure the new concept extraction ability of the models. Specifically, the new concepts extracted by all models are regarded as the overall NCs. Then, the relative recall is calculated as NC \# divided by the number of overall NCs. Accordingly, the relative F1 (R-F1) can also be calculated with Prec. and R-Recall. The results of Table \ref{tb:result0} show that MRC-CE can gain satisfactory precision and recall simultaneously.

\subsubsection{Ablation Study}
We further display the results of ablation study on the Chinese dataset, to investigate the effectiveness of each module in MRC-CE. At first, we took the BERT-MRC with fixed threshold truncation (FTT) as one ablated variant of our framework\cite{BERT-MRC}, denoted as BERT-MRC+FTT. In this variant, we simply chose the spans output by BERT-MRC that have the confidence score higher than 0.8, as the extracted concepts. Furthermore, we respectively appended the rest two modules to BERT-MRC, to propose another two ablated variants, denoted as BERT-MRC+RF and BERT-MRC+RULE.

\begin{table*}[t]
%\vspace{-0.2cm}
%\small
\begin{center}
\caption{CE results of ablation study.}\label{tb:Ablation1}
%\vspace{-0.3cm}
\resizebox{\textwidth}{!}
{
    \begin{tabular}
{|c|c|c|c|c|c|c|}
\hline

\hline
\textbf{Dataset} & \textbf{Model} & \textbf{EC \#} &  \textbf{NC \#} &  \textbf{Prec.} & \textbf{R-Recall} & \textbf{R-F1}\\
\hline

\hline
\multirow{4}{*}{\textbf{CN-DBpedia}} & \footnotesize{BERT-MRC+FTT} & \textbf{522} & 311  & 87.96\% & 83.83\% & 85.84\%\\
\cline{2-7}
& \footnotesize{BERT-MRC+RF} & 519& \textbf{323} & 90.83\% & \textbf{87.06\%} &  88.91\% \\
\cline{2-7}
& \footnotesize{BERT-MRC+RULE} & \textbf{522}& 311 & 91.84\% & 83.83\% & 87.65\% \\
\cline{2-7} 
& \footnotesize{BERT-MRC+RF+RULE} & 519& \textbf{323} & \textbf{92.22\%} & \textbf{87.06\%} & \textbf{89.57\%} \\
\hline

\hline
\multirow{4}{*}{\textbf{Probase}} & \footnotesize{BERT-MRC+FTT} & \textbf{646} & 502  & 79.56\% & 71.21\% & 75.15\%\\
\cline{2-7}
& \footnotesize{BERT-MRC+RF} & 636 & \textbf{626}& 89.44\% & \textbf{88.79\%} & 89.12\%\\
\cline{2-7}
& \footnotesize{BERT-MRC+RULE} & \textbf{646} & 502 & 84.72\% & 71.21\% & 77.38\%\\
\cline{2-7} 
& \footnotesize{BERT-MRC+RF+RULE} & 636 & \textbf{626} & \textbf{90.08\%} & \textbf{88.79\%} & \textbf{89.43\%}\\
\hline

\hline
\end{tabular}
}
\end{center}
\end{table*}
%We randomly selected 500 samples in the test set and get the extraction results from some volunteers. Please note some of indicators mentioned above were too close since the  se model all employ the BERT-MRC module, and thus we did not list the all indicators. 
The comparison results of all variants and MRC-CE (BERT-MRC+RF+RULE) are listed in Table \ref{tb:Ablation1}. The results show that Prec. drops 1.39\% and 0.64\% without RULE (BERT-MRC+RF vs. BERT-MRC+RF+RULE), as well as R-Recall drops 3.23\% and 17.58\% without RF (BERT-MRC+RULE vs. BERT-MRC+RF+RULE) in CN-DBpedia and Probase, respectively. It proves that MRC-CE can obtain better performance with RF and RULE since they are complementary to each other.

Meanwhile, we investigate different BERT-based encoders' influence on CE performance. To this end, we replaced BERT with RoBERTa \cite{RoBERTa} in Concept Extractor module. Then, RF and RULE were also adopted to filter out the candidate spans output by RoBERTa-MRC. The comparison results are shown in Table \ref{tb:Ablation2}, showing that although the variants with RoBERTa gain higher Prec., they are inferior to the indicators with BERT on other metrics. It is possibly due to that, RoBERTa tends to extract the spans the same as the concepts existing in the KG from the texts, and thus it is easy to ignore the new concepts.
\begin{table*}[t]
%\vspace{-0.2cm}
%\small
\begin{center}
\caption{Comparisons between RoBERTa and BERT in CN-DBpedia.}\label{tb:Ablation2}
%\vspace{-0.3cm}
\resizebox{\textwidth}{!}
{
    \begin{tabular}
{|c|c|c|c|c|c|c|c|c|c|}
\hline

\hline
\multirow{2}{*}{\textbf{Encoder}} & \multirow{2}{*}{\textbf{Model}} & \multirow{2}{*}{\textbf{EC\#}} &  \multirow{2}{*}{\textbf{NC\#}} &  \textbf{EC} & \textbf{NC} & \textbf{OC} &  \multirow{2}{*}{\textbf{Prec.}} & \multirow{2}{*}{\textbf{R-Recall.}} & \multirow{2}{*}{\textbf{R-F1.}}\\
& & & &  \textbf{length} & \textbf{Length} & \textbf{ratio} & & &\\
\hline

\hline
\multirow{5}{*}{\textbf{RoBERTa}} & {+FTT} & 
403 & 105  & \multirow{5}{*}{2.63} & 4.97 & 4.75\% & 92.87\% & 25.80\% & 40.38\%\\
\cline{2-4} \cline{6-10}
& {+RF} & 404 & {119} & & 4.92 & 6.69\% & 89.71\% & {29.24\%} & {44.10\%} \\
\cline{2-4} \cline{6-10}
& {+RULE} & {403} & 105 &  & 4.91&  4.73\% & \textbf{95.31\%} & 25.80\% & 40.61\%\\
\cline{2-4} \cline{6-10}
& {+RF+RULE} & {404} & 119 & & \textbf{5.01} & {6.93\%} & {94.92\%} & {29.24\%} & {44.71\%}\\
\hline

\hline
\multirow{5}{*}{\textbf{BERT}} & \footnotesize{+FTT} & \textbf{522} & 311  & \multirow{5}{*}{2.63} & 4.72 & 27.74\% & 87.96\% & 76.41\% & 81.78\%\\
\cline{2-4} \cline{6-10}
& {+RF} & 519 & \textbf{323} & & 4.75& 27.26\% & 90.83\% & \textbf{79.36\%} & 84.71\% \\
\cline{2-4} \cline{6-10}
& {+RULE} & \textbf{522} & 311 & & 4.74 & 27.83\% &  91.84\% & 76.41\% & 83.42\%\\
\cline{2-4} \cline{6-10}
& {+RF+RULE} & {519} & \textbf{323} & & {4.91} &  \textbf{29.35\%} & {92.22\%} & \textbf{79.36\%} & \textbf{85.31\%}\\
\hline

\hline
\end{tabular}
}
\end{center}
\end{table*}

\subsubsection{Case Study}
\setlength{\tabcolsep}{3mm}{
\begin{table*}[t]
%\vspace{-0.2cm}
    \centering
 %\small
    \caption{The CE results of two specific entities along with their existing concepts in the KG.}  \label{tb:case}
    %\vspace{-0.3cm}
\resizebox{\textwidth}{!}
{
    \begin{tabular}{|c|c|c|c|c|}
    \hline
    
\hline
    \textbf{Dataset} & \textbf{Entity} & \textbf{Abstract text} & \textbf{Model} & \textbf{Correct extracted  concept} \\
    \hline
    
  \hline
    & & & Existing concepts & location, station  \\  
     \cline{4-5}
     & &  &  XLNet-NER  & station  \\  
     \cline{4-5}
      & &  Prince station is the  &  FLAIR  & station  \\  
     \cline{4-5}
     CN- & Prince  &  railway station of&  KVMN  & railway station  \\  
     \cline{4-5}
     DBpedia & Station &  JR East Japan  & XLM-R & railway station\\
     \cline{4-5}
     & &  and Tokyo Metro. & BERT-BiLSTM-CRF  & station \\
    \cline{4-5}
    & &  & Hearst Patterns  & the railway station of JR East Japan \\
    \cline{4-5}
     &  & & \multirow{2}{*}{MRC-CE} & station, railway station,\\
     & & & & the railway station of JR East Japan\\
 \hline

 \hline
    & &Franklin Delano Roosevelt& Existing concepts & figure, president, leader, politician \\ 
     \cline{4-5}
  & &   was an American politician &  \multirow{2}{*}{XLNet-NER}  & president,leader  \\  
     & &  who served as the 32nd  &    & figure, politician  \\  
     \cline{4-5}
      &Franklin & president. He became a  &  FLAIR  & politician  \\  
     \cline{4-5}
     &  Delano   & central  figure in world  &  KVMN  & politician  \\
     \cline{4-5}
     Probase  & Roosevelt & events  during the first half & XLM-R & figure, president, leader, politician\\
     \cline{4-5}
     &   & of the 20th century.As a & BERT-BiLSTM-CRF  & politician \\
    \cline{4-5}
    &   & dominant leader of  his party.  & Hearst Patterns  & American politician \\
    \cline{4-5}
     &  & He built New Deal Coalition. & \multirow{2}{*}{MRC-CE} & figure, politician, leader,\\
     & &  &  & president, American politician\\
 \hline
 
\hline
    \end{tabular}
    %\vspace{-0.2cm}
}
\end{table*}
}

We further delve into the concepts extracted by MRC-CE and baselines through some case studies. Table \ref{tb:case} lists the correct concepts extracted by the models along with the existing concepts of two entities\footnote{\emph{Prince Station}'s abstract text and CE results were translated from Chinese.}. It shows that MRC-CE extracts more fine-grained and overlapped concepts from the texts than the baselines.

\section{Applications}
We further demonstrate MRC-CE's advantages on the KG completion of some real applications.
\subsection{KG Completion}
\begin{table*}[t]
%\vspace{-0.1cm}
    \centering
 %\small
    \caption{The statistics of extracted concepts for all entities in CN-DBpedia.} \label{tb:new}
    %\vspace{-0.3cm}
\resizebox{\textwidth}{!}
{
\begin{tabular}{c c c c c c c}
    \toprule
    Data  & instanceOf  & Unique & NC \# & Avg. NC  \#  & Avg. concept \# & Avg. character \#   \\
     type & relation \# &concept \# & &per entity &  per entity & per concept \\
    \midrule
    Original & 11,494,627 & 270,025 & -  & -  &  2.04 & 3.62 \\  
    %\midrule
    Extracted & 9,021,805 & 894,689 & 7,053,900 & 3.16 &  5.20 & 4.92 \\  
    \bottomrule
    \end{tabular}
}
%\vspace{-0.2cm}
\end{table*}
%Based on above experiment results, we ensure that MRC-CE can achieve an excellent concept extraction. Therefore, 
After training MRC-CE, we ran it for all entities in CN-DBpedia to supply substantial new instanceOf relations including unique new concepts. Please note that the instanceOf relations only focus on the relations between the entity and concept. The related statistics of CE results are listed in Table \ref{tb:new}. Please note that the extracted concepts counted in the table include some wrong concepts. According to the table, MRC-CE extracts more than 7,053,900 new instanceOf relations (NC \#). In addition, the extracted concepts have more characters than the original concepts existing in CN-DBpeida. The results justify MRC-CE's capability of extracting large-scale multi-granular concepts for KG completion. 

\subsection{Domain Concept Acquisition}
Our MRC-CE have achieved excellent results concept acquisition on the abstract texts of given entities. 
In order to verify MRC-CE's capability of acquiring concepts for a certain domain, we also employed MRC-CE in the domain of Meituan which is a famous Chinese e-business platform. Specifically, we collected 117,489 Food \& Delight entities in Meituan along with their descriptive texts from CN-DBpedia. After running MRC-CE, we got 458,497 new instanceOf relations from the texts, and the CE precision is 78.0\% based on human assessments on 100 samples, justifying that MRC-CE can be successfully applied to domain concept acquisition.

\section{Conclusion}
In this paper, we propose a concept extraction framework MRC-CE to achieve large-scale multi-granular concept completion for existing KGs. MRC-CE is capable of extracting massive multi-granular concepts from entities' descriptive texts. In our framework, a BERT-based MRC model with a pointer network is built to handle the problems of concept overlap. Meanwhile, a random forest and rule-based pruning are also employed to obtain satisfactory concept extraction (CE) precision and recall simultaneously. Our extensive experiments have justified that our MRC-CE not only has excellent CE performance, but also is competent to acquire large-scale concepts for multilingual KGs. Furthermore, MRC-CE makes a great contribution to supply sufficient concepts for the Chinese KG \emph{CN-Dbpedia}.

%\section*{Acknowledgments}

\bibliographystyle{splncs04}
\bibliography{mybibliography}
\end{document}